\documentclass[twocolumn,prl,aps,showpacs,superscriptaddress]{revtex4-2}
\usepackage{graphicx,graphics}
\usepackage{amsmath}
\usepackage{amssymb}
\usepackage{epstopdf}
\usepackage{amstext}
\usepackage{amsthm}
\usepackage{amsfonts}
\usepackage{latexsym}
\usepackage{array}
\usepackage{xfrac}
\usepackage{lipsum}
\usepackage[toc,page]{appendix}
\usepackage{subfigure}
\newcommand{\igbjd}[1]{}\newcommand{\beqa}{\begin{eqnarray}}
\newcommand{\eeqa}{\end{eqnarray}}
\newcommand{\beq}{\begin{equation}}
\newcommand{\eeq}{\end{equation}}
\usepackage[normalem]{ulem}
\usepackage[colorlinks=true,citecolor=Cerulean,linkcolor=RubineRed,urlcolor=Cerulean,pdftex]{hyperref}
\usepackage{xcolor}
\definecolor{Cerulean}{rgb}{0.,0.59,0.835}
\definecolor{RubineRed}{rgb}{0.61,0.07,0.12}

\newcommand{\1}{{\text A}}
\newcommand{\2}{{\text B}}

\begin{document}
\title{Ultradilute quantum liquid of dipolar atoms in a bilayer}
\author{G. Guijarro}
\author{G. E. Astrakharchik}
\author{J. Boronat}
\affiliation{Departament de F\'isica, Campus Nord B4-B5, 
Universitat Polit\'ecnica de Catalunya,
E-08034 Barcelona, Spain}

\date{February 11, 2022}

\begin{abstract}
We show that ultradilute quantum liquids can be formed with ultracold bosonic dipolar atoms in a bilayer geometry. Contrary to previous realizations of ultradilute liquids, there is no need of stabilizing the system with an additional repulsive short-range potential. The advantage of the proposed system is that dipolar interactions on their own are sufficient for creation of a self-bound state and no additional short-range potential is needed for the stabilization. We perform quantum Monte Carlo simulations and find a rich ground state phase diagram that contains quantum phase transitions between liquid, solid, atomic gas, and molecular gas phases. The stabilization mechanism of the liquid phase is consistent with the microscopic scenario in which the effective dimer-dimer attraction is balanced by an effective three-dimer repulsion. The equilibrium density of the liquid, which is extremely small, can be controlled by the interlayer distance. From the equation of state, we extract the spinodal density, below which the homogeneous system breaks into droplets. Our results offer a new example of a two-dimensional interacting dipolar liquid in a clean and highly controllable setup.
\end{abstract}


\maketitle

\textit{Introduction.} Quantum liquids are self-bound fluids which exhibit quantum mechanical effects at the macroscopic level. The effects of quantum mechanics and quantum statistics, such as the indistinguishability of elementary particles, are crucial in the description of these systems~\cite{leggett2006quantum}. One of the most celebrated examples of quantum liquids is superfluid helium, which played a revolutionary role in the history of quantum physics. The interaction potential between helium atoms features a repulsive short-range hard core due to the action of the Pauli principle acting on the electronic shells and a long-range attractive van der Waals tail due to induced dipole-dipole interaction that tends to hold the atoms together.
A similar interplay between repulsive and attractive interactions can be observed in atomic Bose gases with dipolar interactions under rotation\cite{PhysRevLett.95.200402,PhysRevLett.95.200403}. 

Recently, an entirely new class of quantum liquids has been created in which the quantum fluctuations stabilize the system, which otherwise would be unstable at the mean-field level~\cite{PhysRevLett.115.155302}. A distinguishing feature of such liquids is their ultradilute density, which can be more than eight orders of magnitude lower than that of liquid helium~\cite{Cabrera2017}. So far, two types of ultradilute quantum liquids have been experimentally created: in dipolar systems~\cite{Kadau2016,Schmitt2016,Ferrier2016,Chomaz2016} and in two component Bose-Bose mixtures~\cite{Cabrera2017,Semeghini2018,Ferioli2019}. In both cases, distinct kinds of interaction potentials were needed. That is, in dipolar systems an additional repulsive potential was required to stabilize the system. In mixtures, the attractive interaction between different species~(AB) was balanced by repulsive same-species interactions (AA and BB) which resulted in three different types of interactions with the complication of being impossible to tune them independently with only one experimental parameter (magnetic field). As a result, in all previous experiments it was necessary to fine-tune two or three types of interactions. While both kinds of systems allow the creation of ultradilute quantum liquids, it is yet an open question if just a single kind of physical interaction might be sufficient.

We argue that quantum dipolar bosons in a bilayer geometry may serve as a simpler and cleaner system in which there is no need of superimposing short-range interactions. If the dipolar moments of the bosons are oriented perpendicularly to the parallel layers, there is a competing effect between repulsive intralayer and partially attractive interlayer interactions, which can produce interesting few-and many-body states. 
The interlayer attractive potential energy is dominated by the dimer contribution, which is a function of the interlayer distance, and has a much weaker density dependence as opposed to the intralayer repulsive potential energy.
As a result, the attractive part of the interlayer interaction potential can overcome the repulsive one which eventually induces a phase transition from atomic to pair superfluids, as discussed in Refs.~\cite{Macia2014,PhysRevA.94.063630}.
Recently, it has been predicted that addition of a three-dimer repulsion to a two-dimer attraction could stabilize a many-body liquid~\cite{PhysRevA.101.041602}. It is therefore an open challenge to determine the existence, formation mechanism, and properties of the self-bound many-body dipolar system in the bilayer geometry.

In this Letter, we study a two-dimensional system of dipolar bosons confined to a bilayer setup. We calculate the ground-state phase diagram as a function of the density and the separation between layers by using exact quantum Monte Carlo methods. The key result of our work is the prediction of a homogeneous liquid in this system. The liquid is stable in a wide range of densities and interlayer values, including experimentally feasible ones. We find that the critical interlayer separation at which the liquid to gas transition happens is the same as the threshold value at which the effective interaction between dimers changes from attractive to repulsive. We characterize the liquid by calculating its equation of state, the condensate fraction, and the equilibrium and spinodal densities.

\begin{figure*}[htp]
\centering
\subfigure{\includegraphics[width=0.48\textwidth]{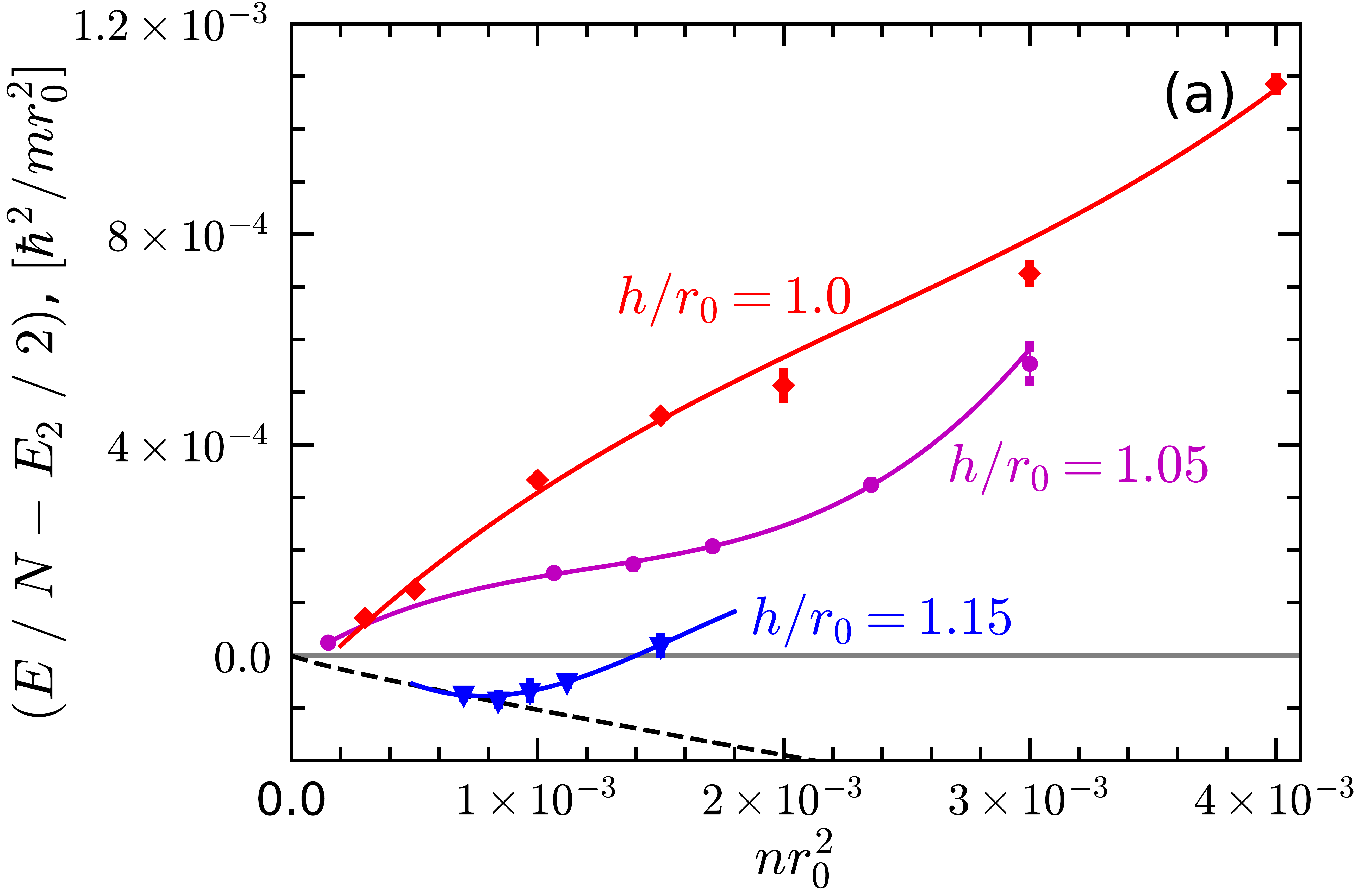}}\quad 
\subfigure{\includegraphics[width=0.48\textwidth]{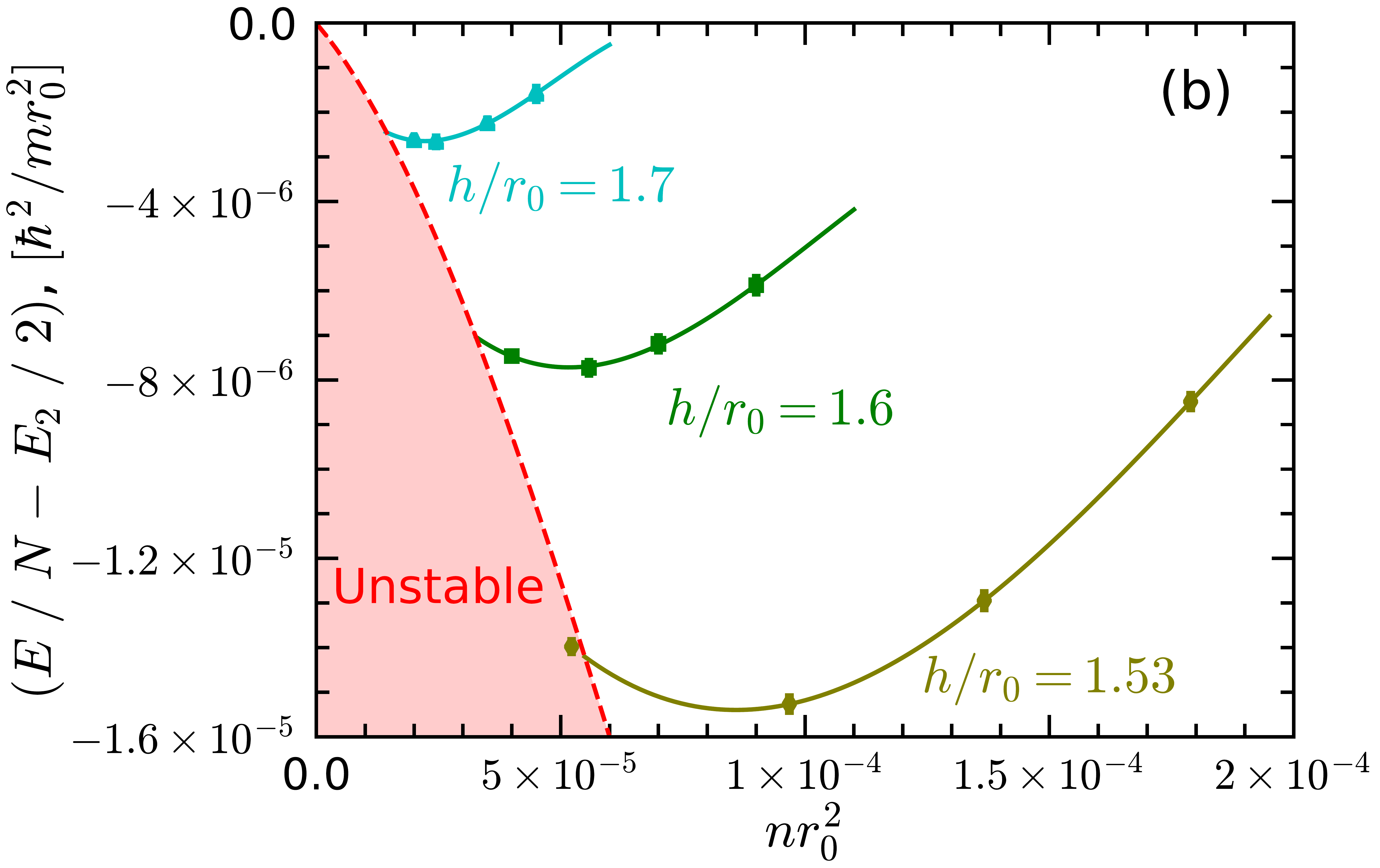}} 
\caption{Energy per particle $E/N$ with one-half of the dimer binding energy $E_2/2$ subtracted as a function of the total density $nr_0^2$ for different values of the interlayer distance $h/r_0$. The solid curves show polynomial fits to the equations of state. Panel (a): From top to bottom the results correspond to the equations of state of a gas ($h/r_0=1.0$, $h/r_0=1.05$), and liquid ($h/r_0=1.15$) phases. The dashed line corresponds to the mean-field approximation for an attractive molecular gas $E/N=-\pi\hbar^2n/4m\text{ln}[na^2_{dd}]$, where $a_{dd}$ is the dimer-dimer scattering length. Panel (b): The three curves correspond to the equation of state of a liquid phase. The region where the liquid is unstable is shown as a shaded (red) area. Its boundary, delimited by a dashed curve, is defined by the spinodal points.}
\label{Fig:EquationOfState}
\end{figure*}

\textit{Hamiltonian.} We consider $N$ bosons of mass $m$ and dipole moment $d$ confined to two parallel layers separated by a distance $h$. It is assumed that the dipolar moment of each boson is aligned perpendicularly to the planes by an external field. Also, we suppose that the confinement to each plane is so tight that there is no interlayer tunneling and that transverse degrees of freedom are frozen. The Hamiltonian of this system is given by
\begin{equation}
\begin{aligned}
\label{Hamiltonian}
H=&-\frac{\hbar^2}{2m}\sum_{i=1}^{N_{\1}} \nabla^2_i-
\frac{\hbar^2}{2m}\sum_{\alpha}^{N_{\2}} \nabla_\alpha^2\\
+&\sum_{i<j}^{N_{\1}} \frac{d^2}{r^3_{ij}}+\sum_{\alpha<\beta}^{N_{\2}}
\frac{d^2}{r^3_{\alpha\beta}}+\sum_{i,\alpha}^{N_{\1},N_{\2}}
\frac{d^2(r_{i\alpha}^2-2h^2)}{(r_{i\alpha}^2+h^2)^{5/2}},
\end{aligned}
\end{equation} 
where Latin (Greek) indices run over each of $N_\1$ ($N_\2$) dipoles in the top (bottom) layer. The first two terms in the Hamiltonian~(\ref{Hamiltonian}) correspond to the kinetic energy and the next two terms are the intralayer dipolar interaction, which is always repulsive and falls off with a power law $1/r^3$. The last term describes the interlayer potential, which is attractive at short distances and repulsive for large values of $r$, where $r$ is the in-plane distance between dipoles. The interlayer potential always supports at least one bound (dimer) state. Its binding energy $E_2$ diverges when $h\to 0$ and exponentially vanishes in the limit of large interlayer separation~\cite{Shih2009,Armstrong2010,Klawunn2010,Baranov2011}. The dipolar length $r_0=md^2/\hbar^2$ is used as a unit of length.

\textit{Method.} We have studied the ground-state properties of the dipolar system using the diffusion Monte Carlo (DMC) method~\cite{BoronatCasulleras1994}. The DMC algorithm solves in a stochastic way the many-body imaginary-time Schr\"odinger equation and is based directly on the microscopic Hamiltonian~(\ref{Hamiltonian}). In this way, the DMC method allows us to calculate the exact ground-state energy of the system, as well as other properties, within controllable statistical errors. 

As usual in DMC calculations, we employ a guiding wave function for importance sampling to reduce the variance to a manageable level. In this work, we use two guiding trial wave functions: the first one is of Jastrow form, composed as a pair product of three different types of two-body correlation terms, 
\begin{equation}
\label{eq2}
\Psi_J(\mathbf{r}_1,\dots,\mathbf{r}_N)=\prod_{i<j}^{N_\1} f_{\1\1}(r_{ij})
\prod_{\alpha<\beta}^{N_\2} f_{\2\2}(r_{\alpha\beta})
\prod_{i,\alpha}^{N_\1,N_\2} f_{\1\2}(r_{i\alpha}) \ ,
\nonumber
\end{equation}
while the second one explicitly takes into account a possible formation of AB dimers with an appropriate symmetrization,
\begin{equation}
\begin{aligned}
\label{eq03}
\Psi_S(\mathbf{r}_1,\dots,\mathbf{r}_N)=&\prod_{i<j}^{N_\1}f_{\1\1}(r_{ij})
\prod_{\alpha<\beta}^{N_\2}f_{\2\2}(r_{\alpha\beta})\\
\times&\Bigg[\prod_{i=1}^{N_\1}\sum_{\alpha=1}^{N_\2}f_{\1\2}(r_{i\alpha})+
\prod_{\alpha=1}^{N_\2}\sum_{i=1}^{N_\1}f_{\1\2}(r_{i\alpha})\Bigg].
\nonumber
\end{aligned}
\end{equation}
While the variance depends on a specific choice of the guiding wave function both choices result in the same DMC energy within statistical errors validating the consistency of the method.

Intraspecies correlations at short distances, $r<R_0$, are modeled by the zero-energy two-body scattering solution $f_{\1\1}(r)=f_{\2\2}(r)=C_0 K_0(2\sqrt{r_0/r})$, with $K_0(r)$ the modified Bessel function and $R_0$ a variational parameter~\cite{PhysRevLett.98.060405}. For distances larger than $R_0$ we choose $f_{\1\1}(r)=f_{\2\2}(r)=C_1 \text{exp}[ -\frac{C_2}{r} - \frac{C_2}{L-r}]$ which describes two-dimensional phonons~\cite{PhysRev.155.88}, $L$ being the simulation box length. The constants $C_0$, $C_1$ and $C_2$ are fixed by imposing continuity of the function and its first derivative at the matching distance $R_0$, and also that $f_{\1\1}(L/2)=1$. The interspecies correlations are described by the dimer wave function $f_{\1\2}(r)$ up to $R_1$. Then, we impose the boundary conditions $f^{'}_{\1\2}(R_1)=0$ and $f_{\1\2}(r)=1$ for $0<R_1<L/2$.

For simplicity, we assume a population-balanced system $N_\1=N_\2=N/2$ where $N$ is the total number of dipoles. In order to approximate the properties of the extended system, we perform DMC simulations in a square box with side length $L$ and impose periodic boundary conditions. The total density of the system is defined as $n=N/L^2$. 

The dipolar potential is a quasi-long ranged one in two dimensions, therefore its truncation at $L/2$ produces significant finite-size corrections. 
The average energy $E_{int}$ of the interaction potential $V(r)$ which is a two-body operator, can be expressed in terms of the pair distribution function $g(r)$ as $E_{int}/N = 1/(2n)\int_0^\infty V(r) g(r) d{\bf r}$. 
Applied to the bilayer geometry, the finite-size effects can be significantly diminished by adding the tail energy,
\begin{equation}
\begin{aligned}
\frac{E_{{\rm tail}}(n,L)}{L^2}=\int_{L/2}^{\infty}& \bigg[ 
\frac{1}{2}\frac{d^2}{r^3}g_{\1\1}(r)+
\frac{1}{2}\frac{d^2}{r^3}g_{\2\2}(r)+\bigg.\\&\bigg. 
\frac{d^2(r^2-2h^2)}{(r^2+h^2)^{5/2}}g_{\1\2}(r) \bigg]2\pi rdr .
\label{eq5}
\end{aligned}
\end{equation}

We denote by $g_{\1\1}(r)=g_{\2\2}(r)$ and $g_{\1\2}(r)$ the intra- and interspecies pair distribution functions. An approximate value of the tail energy~(\ref{eq5}) is obtained by ignoring correlations at large distances, $g_{\1\1}(r)\to n_{\1}^2$, $g_{\2\2}(r)\to n_{\2}^2$ and $g_{\1\2}(r)\to n_{\1}n_{\2}$, which leads to
\begin{equation}
\frac{E_{{\rm tail}}}{N}=\frac{\pi d^2 n^{3/2}}{\sqrt{N}}+
\frac{\pi d^2N}{(4h^2+N/n)^{3/2}}.
\label{eq6}
\end{equation}
Addition of the tail energy~(\ref{eq6}) to the DMC data allows to significantly reduce the finite-size dependence. In the end, we perform extrapolation to the thermodynamic limit using the law $E(N)=E_{th}+C/N^{1/2}$, with $C$ a fitting parameter and report the obtained energy~\cite{SM}.
\begin{figure}[htp]
\centering
\includegraphics[width=0.5\textwidth]{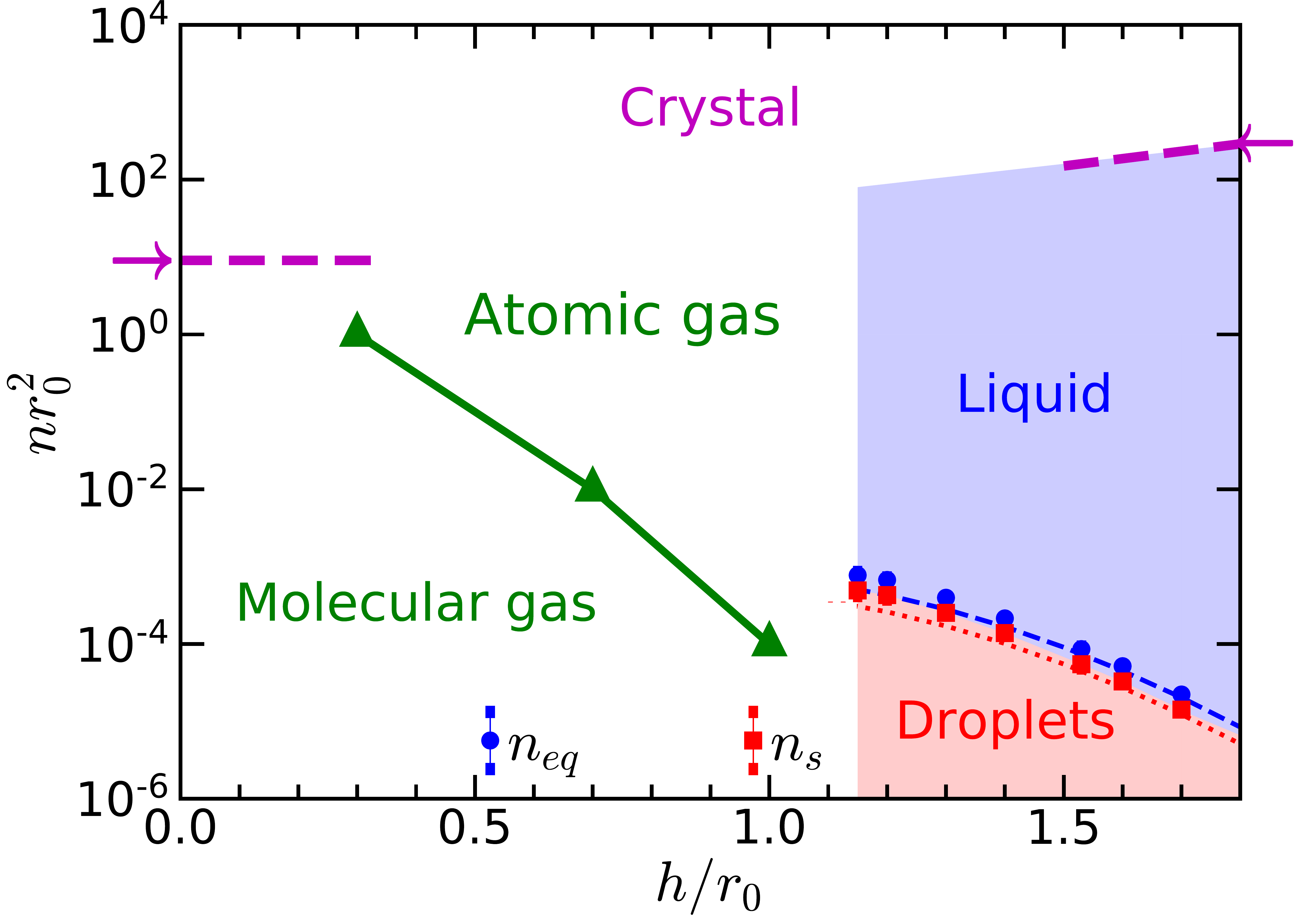}
\caption{Ground-state phase diagram as a function of the total density $nr_0^2$ and the interlayer distance $h/r_0$. The green triangles correspond to the transition points between an atomic and a molecular gas~\cite{Macia2014}. The two arrows show the critical density of crystallization of a single-layer of particles (right) and of molecules (left)~\cite{PhysRevLett.98.060405}. The liquid and droplet phases appear for $h/r_0> 1.1$, blue circles correspond to the equilibrium densities $n_{eq}$ of the liquid and red squares correspond to the spinodal densities $n_s$. Curves correspond to the Bogoliubov approximation for a 2D Bose-Bose mixture with attractive interspecies and repulsive intraspecies short-range interactions~\cite{Petrov2016}.
}
\label{Fig:PhaseDiagram}
\end{figure}

\textit{Results.} The existence of a gas or a liquid phase can be inferred from the equation of state. The dependence of the energy on the total density is reported in Fig.~\ref{Fig:EquationOfState} for characteristic values of the interlayer distance. We observe that for $h/r_0=1.0$, the energy per particle monotonically increases with the density. The smallest energy (zero) is obtained at vanishing density and thus this corresponds to a gas phase. By slightly increasing the interlayer distance to $h/r_0=1.05$, we find that the energy as a function of density shows a tiny inflection point. This could suggest the presence of a first-order gas-liquid phase transition, so that the gas at low density coexists with a liquid at higher density. However, we have not been able to build a double tangent Maxwell construction to characterize this transition because of the minute change observed in the curvature of the equation of state. Drastically different behavior is observed as the interlayer separation is further increased. That is, the energy per particle becomes negative and develops a minimum at a finite density for $h/r_0\ge1.1$. The position of the minimum corresponds to the equilibrium density. The existence of this minimum proves the stability of a homogeneous liquid phase. A possible underlying microscopic mechanism for the stabilization of the liquid is due to effective three-dimer repulsion which balances an effective dimer-dimer attraction as proposed in Ref.~\cite{PhysRevA.101.041602}. Indeed, the three-dimer repulsive contribution is negligible for small densities and we find an energy which closely follows that of an attractive gas of dimers, shown by a dashed line in Fig.~\ref{Fig:EquationOfState}a. We found that the interlayer critical value for the liquid to gas transition ($h/r_0\approx 1.1$) is the same as the threshold value for the four-body bound state of dipolar bosons, when the tetramer breaks into two dimers~\cite{PhysRevA.101.041602}. The non-monotonic dependence of the binding energy on $h/r_0$ is shared between the few-body AABB, AAABBB, etc clusters\cite{PhysRevA.101.041602} and the thermodynamic liquid.

The equations of state are used to extract the equilibrium $n_{eq}$ and spinodal $n_s$ densities, which are defined by the conditions $\frac{\partial E/N}{\partial n}=0$ and $\frac{\partial P}{\partial n}=0$, respectively with $P=n^2 \frac{\partial E/N}{\partial n}$ being the pressure. The resulting ground-state phase diagram is reported in Fig.~\ref{Fig:PhaseDiagram} as a function of the total density $nr_0^2$ and of the interlayer distance $h/r_0$. The self-bound many-body phases are formed for large interlayer separations, $h/r_0>1.1$. Below the spinodal curve (dotted line) the homogeneous liquid is unstable with respect to droplet formation. The stable liquid phase appears above the spinodal curve. Remarkably, this phase exists in a wide range of densities and interlayer values. The equilibrium density (dashed line) can be adjusted by changing the separation between the layers: $n_{eq}$ decreases as $h/r_0$ increases. For large separations $h$, the liquid becomes very dilute and, in this weakly-interacting regime, it is possible to make a comparison with the predictions of Bogoliubov theory~\cite{Petrov2016} developed for short-range potentials.
The best agreement is found for the smallest equilibrium and spinodal densities, i.e., for the largest $h$, for which the dipolar potential is well approximated by a contact potential with the same $s$-wave scattering length. The gaseous and self-bound phases are separated by the threshold $h/r_0\approx1.1$ at which the effective dimer-dimer interaction changes its sign~\cite{PhysRevA.101.041602} from repulsion (gas) to attraction (homogeneous liquid or droplets). The gaseous regime features a second-order phase transition between atomic and molecular gas phases which on a qualitative level occurs when the molecular binding energy approaches the chemical potential. In the molecular gas phase, the atomic condensate is absent while the molecular one is finite~\cite{Macia2014}. On the other hand, in the atomic gas one observes an atomic condensate and the system features a strong Andreev-Bashkin drag between superfluids in different layers~\cite{Nespolo_2017}. The gas phase is characterized by a quantum phase transition from a molecular to an atomic superfluid as the interlayer distance increases. Indeed, we have
verified that an atomic condensate is present in the homogeneous liquid.

As the density of the liquid is increased, the potential energy starts to dominate and eventually a triangular crystal is formed. For large separation between layers, two independent atomic crystals are formed and the phase transition occurs when the density per layer reaches the same critical value as in a single-layer geometry, $nr_0^2\approx 290$~\cite{PhysRevLett.98.060405,Cinti2017}. In the limit of small interlayer separations, a single molecular crystal is formed at density $nr_0^2\approx 9$.

\begin{figure}
\centering
\includegraphics[width=0.5\textwidth]{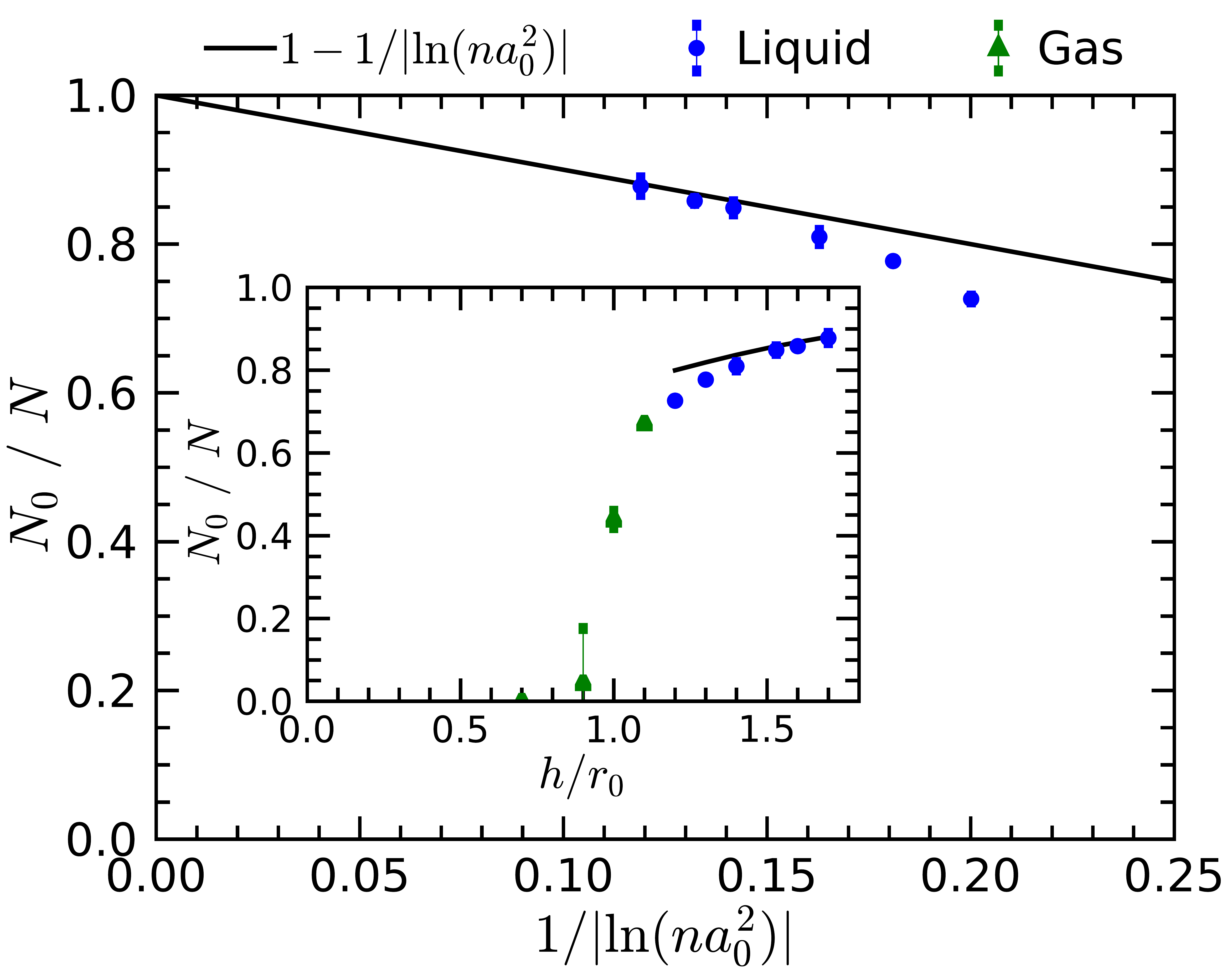} 
\caption{Depletion of the condensate fraction $N_0/N$ vs $1/|\ln(na_0^2)|$. The blue circles correspond to the liquid phase at the equilibrium density. The solid line corresponds to the quantum depletion of short-range potentials having $s$-wave scattering length $a_0$ in two dimensions $1-1/|\ln(na_0^2)|$. Inset: $N_0/N$ as a function of the interlayer distance $h/r_0$, the green triangles correspond to the gas phase at total density $nr_0^2=10^{-3}$.}
\label{Fig:CondensateFraction}
\end{figure}

In order to quantify the quantum coherence in the system, we have calculated the atomic condensate fraction $N_0/N $ as the off-diagonal long-range limit $|\mathbf{r}-\mathbf{r'}|\to \infty$ of the one-body density matrix (OBDM) $n^{(1)}(\mathbf{r},\mathbf{r'})=\langle \hat{\Psi}^{\dagger}(\mathbf{r})\hat{\Psi}(\mathbf{r'}) \rangle$, where $\hat{\Psi}^{\dagger}(\mathbf{r}) (\hat{\Psi}(\mathbf{r}))$ is the field operator that creates (annihilates) a particle at the point $\mathbf{r}$~\cite{LIFSHITZ198085}. In Fig.~\ref{Fig:CondensateFraction}, we report the condensate fraction $N_0/N$ as a function of $1/|\ln(na_0^2)|$, for the dipolar liquid at the equilibrium density, where $a_0={\rm e}^{2\gamma}r_0$ is the $s$-wave scattering length and $\gamma\approx0.577$ is the Euler constant. In the very dilute limit, we find a good agreement with the quantum depletion $1/|\ln(na_0^2)|$ calculated in Bogoliubov theory for short-range potentials. The equilibrium density has a strong dependence on the interlayer separation $h$ (see Fig.~\ref{Fig:PhaseDiagram}). For liquids formed at separations $h\gtrsim 1.6$ the perturbative result is expected to hold. In the inset of Fig.~\ref{Fig:CondensateFraction} we report the condensate fraction as a function of the interlayer separation $h$. The liquid exists for large separations between the layers $h$. As $h$ is decreased the equilibrium density grows up until it reaches $nr_0^2\approx 10^{-3}$ at $h/r_0\approx 1.1$ where a phase transition from a liquid to a gas happens. For smaller separations, the liquid does not exist and we show the condensate fraction in the gas with the density fixed to $nr_0^2=10^{-3}$. The condensate fraction rapidly drops to zero signaling a phase transition from atomic to molecular gas.

\textit{Possible parameters for experimental realization.} Our results open new perspectives for experimental observation of quantum liquids in quasi-two-dimensional geometries. The predicted liquid, for ratios $h/r_0>1.1$, can be realized by using bosonic dipolar molecules produced with mixtures of $^{87}$Rb$^{133}$Cs~\cite{PhysRevLett.113.205301,PhysRevLett.113.255301} and $^{23}$Na$^{87}$Rb~\cite{PhysRevLett.116.205303,PhysRevA.97.020501} characterized by dipolar lengths $r_0\sim5\times 10^{-6}$m and $2\times10^{-5}$m, respectively. Magnetic dipolar $^{164}$Dy$_2$~\cite{PhysRevLett.107.190401} and $^{168}$Er$^{164}$Dy~\cite{PhysRevLett.121.213601} ($r_0\sim2\times 10^{-7}$m and $r_0\sim1\times 10^{-7}$m, respectively) molecules can also be used. The interlayer distance, one-half of the laser wavelength $\lambda/2$, has typical values of $h\approx (2-5)\times 10^{-7}$m. In an experimental realization the bilayer has a quasi-two-dimensional geometry, that is, each layer has a finite width. The effective intralayer interaction for a one layer of dipolar atoms with finite width was derived in Ref.~\cite{PhysRevLett.105.255301}. Considering a Gaussian profile in the transverse direction to the planes, the length scale of the transverse confinement $a_\perp=\frac{\lambda}{2\pi}s^{-1/4}$~\cite{HADZIBABIC201295} is related to the laser beam wavelength $\lambda$ and its strength, which commonly is quantified as the height $s$ of the optical lattice in units of the recoil energy. A typical value of $s=16$ results in $a_\perp = 0.08\lambda$, that is the transverse size can be significantly smaller than the distance between layers. For these typical experimental parameters, we have found no significant differences between a quasi-two and a two-dimensional repulsive potential. For attractive interactions, we have found that the dimer energy calculated with our model and with a quasi-two-dimensional model~\cite{PhysRevA.81.063616}
differs at most by 20\%. Therefore, we conclude that the effects of considering a quasi-two-dimensional model do not change our main conclusions.

\textit{Conclusions.} In conclusion, by using exact QMC methods we have shown that it is feasible to create ultradilute quantum liquids in atomic systems interacting with purely dipolar interactions (i.e., no $s$-wave resonance is needed) when confined to a bilayer geometry. The stabilization mechanism is consistent with a microscopic description in which the liquid state is formed from the balance of a dimer-dimer attraction and an effective three-dimer repulsion. A dipolar bilayer possesses a rich phase diagram with quantum phase transitions between gas, solid phases (known before), and a liquid phase (newly predicted). From the equations of state, we extracted the spinodal and equilibrium densities, which are controllable through the interlayer distance. The equilibrium density decreases as the interlayer distance increases, allowing access to ultra-dilute liquids in a stable setup. Remarkably, the liquid state exists in a wide range of densities and interlayer separations which are experimentally accessible. Therefore, our results offer a new example of an ultradilute quantum liquid which can be experimentally realized in a clean and highly controllable setup.

\begin{acknowledgments}
\textit{Acknowledgments}. This work has been supported by the Ministerio de Economia, Industria y Competitividad (MINECO, Spain) under grant No.~FIS2017-84114-C2-1-P. We acknowledge financial support from Secretaria d'Universitats i Recerca del Departament d'Empresa i Coneixement de la Generalitat de Catalunya, co-funded by the European Union Regional Development Fund within the ERDF Operational Program of Catalunya (project QuantumCat, ref.~001-P-001644). The authors thankfully acknowledge the computer resources at Cibeles and the technical support provided by Barcelona Supercomputing Center (RES-FI-2021-1-0020). G.G. acknowledges a fellowship by CONACYT (Mexico).
\end{acknowledgments}

\section{Supplementary materials}


\subsection{Finite size effects}

After adding the tail energy $E_{\rm tail}$ to the DMC energy $E_{\rm DMC}$, we extrapolate the energy $E(N)=E_{\rm DMC}+E_{\rm tail}$ to the thermodynamic limit value $E_{\rm th}$ using the fitting formula
\begin{equation}
    E(N)=E_{\rm th}+\frac{C}{\sqrt{N}}, 
\label{Eq:6.10}
\end{equation}
where $C$ is a fitting parameter.

\begin{figure}[h]
	\centering
    \includegraphics[width=0.5\textwidth]{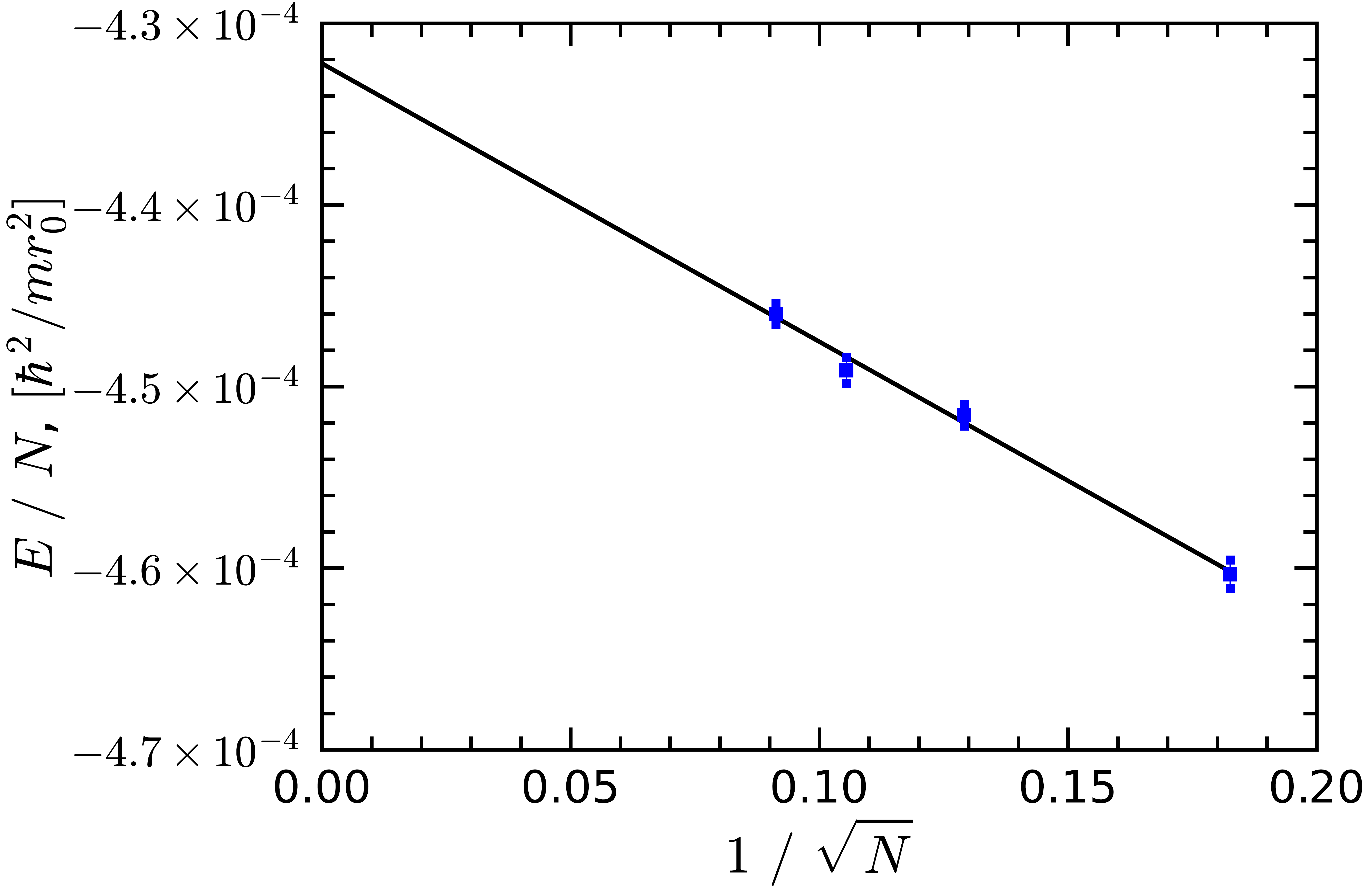}
    \caption{An example of the finite-size dependence for the energy
        in the liquid phase at the dimensionless density $nr^2_0=0.001033$
        and $h/r_0=1.2$.
        Symbols, DMC energy (with added the tail energy);
        solid line, fit $E_{\rm th}+C/\sqrt{N}$.}
\label{Fig:FiniteSize}
\end{figure}

In Fig.~\ref{Fig:FiniteSize}, we show an example of the finite-size dependence of the ground-state energy.
In it, we consider a liquid phase with density $nr_0^2=0.001033$ and interlayer distance $h/r_0=1.2$.
We observe that the energy dependence on the number of particles scales as
$1/\sqrt{N}$, contrary to the law $1/N$, typical for fast decaying potentials. 
We find that fitting function~(\ref{Eq:6.10}) describes well the finite-size dependence. 

The number of particles used in this study ranges from $N=30$ up to $N=120$. All the energies reported in our work are corrected to the thermodynamic limit using this functional law. The described procedure is used in gas and liquid phases.

\subsection{Polarization}

The pairing properties of the different phases present in the system can be further characterized by calculating the ground-state energy dependence on small polarization. This dependence can be linear or quadratic depending on the molecular or atomic nature of the system, respectively. This can be obtained by slightly imbalanced the number of particles in the bottom $N_\1$ and top $N_\2$ layers, while keeping fixed the total number of particles $N_\1+N_\2$. The polarization is defined as
\begin{equation}
    P=\frac{N_\1-N_\2}{N_\1+N_\2},
\end{equation}
and small is small in a slightly unbalanced system, $|P|\ll 1$.

For an atomic condensate, the ground-state energy dependence on small polarization is quadratic
\begin{equation}
    E(P)=E(0)+N(n/2\chi_s)P^2,
\end{equation}
where $E(0)$ is the ground-state energy of the balanced system and $\chi_s$ is the spin susceptibility associated with the dispersion of spin waves of the magnetization density $n_t-n_d$ with speed of sound $c_s=\sqrt{n/m\chi_s}$. In this case the low-lying excitations are coupled phonon modes of the two layers.
\begin{figure}[t!]
	\centering
    \includegraphics[width=0.5\textwidth]{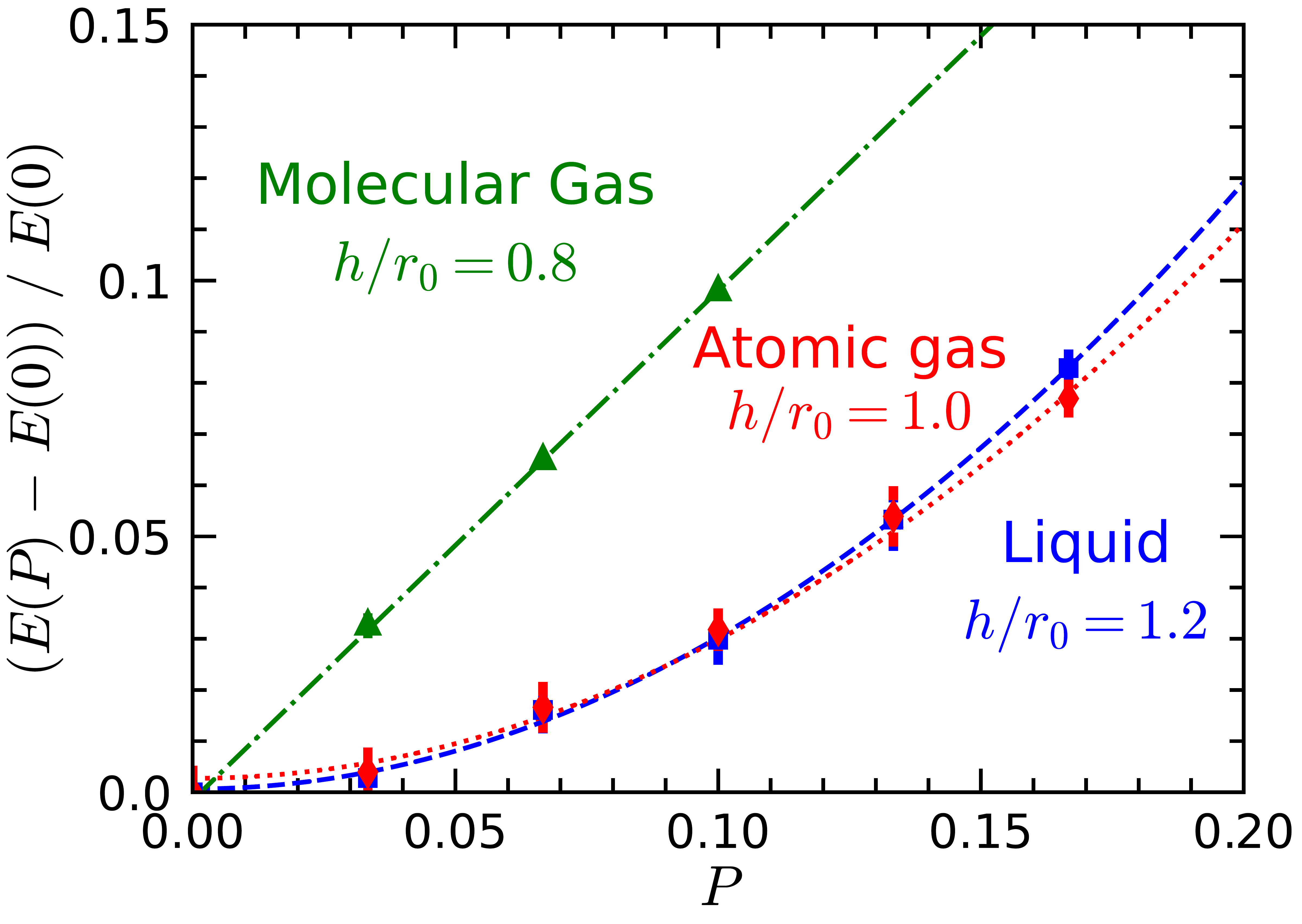}
    \caption{Ground-state energy $E(P)$ as a function of the polarization         $P$ for three values of $h/r_0$ in the molecular gas ($nr_0^2=0.001$),
        atomic gas ($nr_0^2=0.001$),
        and liquid phase (equilibrium density).}
    \label{Fig:Polarization}
\end{figure}

For a molecular superfluid phase the ground-state energy is a linear function of the polarization 
\begin{equation}
    E(P)=E(0)+N\Delta_{\rm gap}P,
\end{equation}
in this case an energy $\Delta_{\rm gap}$ is needed to break a pair and spin excitations are gapped.

Examples of the different behaviors of $E(P)$ are reported in Fig.~\ref{Fig:Polarization} for three values of $h/r_0$ corresponding to the molecular gas, atomic gas, and liquid phases. We notice that $E(P)$ is a quadratic function of $P$ for the liquid state, therefore the liquid is a liquid of atoms and not a liquid of molecules or dimers. 

%

\end{document}